\documentclass[prl,aps,twocolumn]{revtex4-1}

\usepackage[pdftex]{graphicx}
\usepackage{amsbsy,amssymb,amsmath,bm,mathtools}

\usepackage{xspace}
\usepackage{bm}
\usepackage{color}
 
\usepackage{url}
\usepackage[section]{placeins}
\usepackage[colorlinks=true,linkcolor=blue,citecolor=blue]{hyperref}

\begin{document}


\title{Arrested phase separation in double-exchange models: \\ machine-learning enabled large-scale simulation}

\author{Puhan Zhang}


\author{Gia-Wei Chern}
\affiliation{Department of Physics, University of Virginia, Charlottesville, VA 22904, USA}

\date{\today}

\begin{abstract}
We present large-scale dynamical simulations of electronic phase separation in the single-band double-exchange model based on deep-learning neural-network potentials trained from small-size exact diagonalization solutions. 
We uncover an intriguing correlation-induced freezing behavior as doped holes are segregated from half-filled insulating background during equilibration. While the aggregation of holes is stabilized by the formation of ferromagnetic clusters through Hund's coupling between charge carriers and local magnetic moments, this stabilization also creates confining potentials for holes when antiferromagnetic spin-spin correlation is well developed in the background. The dramatically reduced mobility of the self-trapped holes prematurely disrupts further growth of the ferromagnetic clusters, leading to an arrested phase separation. Implications of our findings for phase separation dynamics in materials that exhibit colossal magnetoresistance effect are  discussed.
\end{abstract}

\maketitle

The subject of phase separation dynamics is of significant importance in many branches of physics, materials science, and biology~\cite{gunton83,puri09,cross12,onuki02,bray94}. Dynamically, phase separation occurs when a homogeneous system is placed in an out-of-equilibrium state due to a rapid change in thermodynamic variables such as temperature. The system then evolves toward an inhomogeneous state of coexisting phases. This intrinsically non-equilibrium and nonlinear process involves the formation, growth, and coarsening of domains of ordered phases. Substantial progress has been made in understanding the phase-separation kinetics over the past few decades. In particular, it has been shown that phase separation at late times exhibits a dynamical scaling and is controlled by a characteristic length scale $L$ which follows a power-law $L(t) \sim t^{\alpha}$, where the growth-exponent $\alpha$ depends mainly on dimensionality and conservation of the order-parameter.

Phase separation also plays a crucial role in the functionality of strongly correlated electron systems~\cite{schulz89,emery90,white00,tranquada95,yee15,kivelson03,dagotto_book,dagotto05,moreo99,dagotto01,mathur03,nagaev02}.  
A~case in point is the complex inhomogeneous states observed in manganites and magnetic semiconductors that exhibit the colossal magnetoresistance (CMR) effect~\cite{dagotto_book,dagotto05,moreo99,dagotto01,mathur03,nagaev02}. These nanoscale textures arise from the segregation of hole-rich ferromagnetic clusters from the half-filled antiferromagnetic  domains~\cite{fath99,renner02,salamon01}. An intriguing scenario for CMR is the field-induced percolating transition of metallic nano-clusters in such a phase-separated state~\cite{uehara99,zhang02}.  
Since the number of doped carriers is conserved, the segregation process of such conserved field was first studied in the classic works of Lifshitz, Slyozov~\cite{lifshitz61}, and Wagner~\cite{wagner61} (LSW), who predicted a growth exponent of~$\alpha = 1/3$.

Despite extensive works on properties of mixed-phase states in CMR materials, the kinetics of phase separation driven by electron-correlation has yet to be investigated. Important questions, such as whether the phase separation exhibits dynamical scaling and does the late-stage domain growth indeed follow the LSW power law, remain unanswered. 
On the theoretical side, the lack of progress is partially due to the difficulty in performing large-scale dynamical simulations of electronic phase separation. While several numerical techniques, such as molecular dynamics and phase-field method~\cite{collins86,valls90,steinbach13}, have been developed to simulate pattern formation in material systems such as binary alloys or polymers, conventional approaches often rely on empirical energy models and cannot describe the intricate electron correlation effects. A comprehensive modeling of correlation-induced phase separation requires properly incorporating microscopic electronic processes into mesoscopic spatial-temporal pattern dynamics. Yet, such multi-scale approaches are limited to small systems due to the expensive repeated electronic structure calculations.

In this paper, we overcome the difficulties of multi-scale modeling by applying machine learning (ML) methods to enable large-scale simulations of phase separation phenomena in the double-exchange (DE) model~\cite{zener51,anderson55,degennes60}, which plays a center role in the modeling of CMR materials. The central idea of our approach is to develop deep-learning neural networks (NN) that emulate the time-consuming exact diagonalization required for computing the exchange forces on spins. In this respect, the NN can be viewed as a complex empirical potential model, which offers the accuracy of quantum calculations. 
We consider the single-band DE model on a square lattice,
\begin{eqnarray}
	\label{eq:H_DE}
	\hat{\mathcal{H}} = -t_{\rm nn} \sum_{\langle ij \rangle} \left( \hat{c}^{\dagger}_{i \alpha} \hat{c}^{\;}_{j \alpha} + {\rm h.c.} \right)
	- J_H \sum_{i} \mathbf S_i \cdot \hat{c}^{\dagger}_{i\alpha}  {\bm{\sigma}_{\alpha\beta}} \hat{c}^{\;}_{i\beta}, \qquad
\end{eqnarray}
where $\hat{c}^\dagger_{i \alpha}/\hat{c}_{i, \alpha}$ are creation/annihilation operators of electron with spin $\alpha = \uparrow, \downarrow$ at site $i$, repeated indices $\alpha, \beta$ implies spin summation,  $ \langle ij \rangle$ indicates the nearest neighbors (NN), $t_{\rm nn}$ is the NN electron hopping constant, $J_H$ is the local Hund's rule coupling between electron spin and local magnetic moment $\mathbf S_i$, which are assumed to be classical spins of length $S = 1$.  The square-lattice DE model has been extensively studied theoretically~\cite{varma96,yunoki98,dagotto98,chattopadhyay01,pekker05}. In particular, when the system is slightly hole-doped from half-filling, a mixed-phase state consisting of hole-rich ferromagnetic puddles embedded in the half-filled antiferromagnetic insulator emerges as a stable thermodynamic phase at strong Hund's coupling~\cite{yunoki98,dagotto98,chattopadhyay01}.

\begin{figure}
\includegraphics[width=1.0\columnwidth]{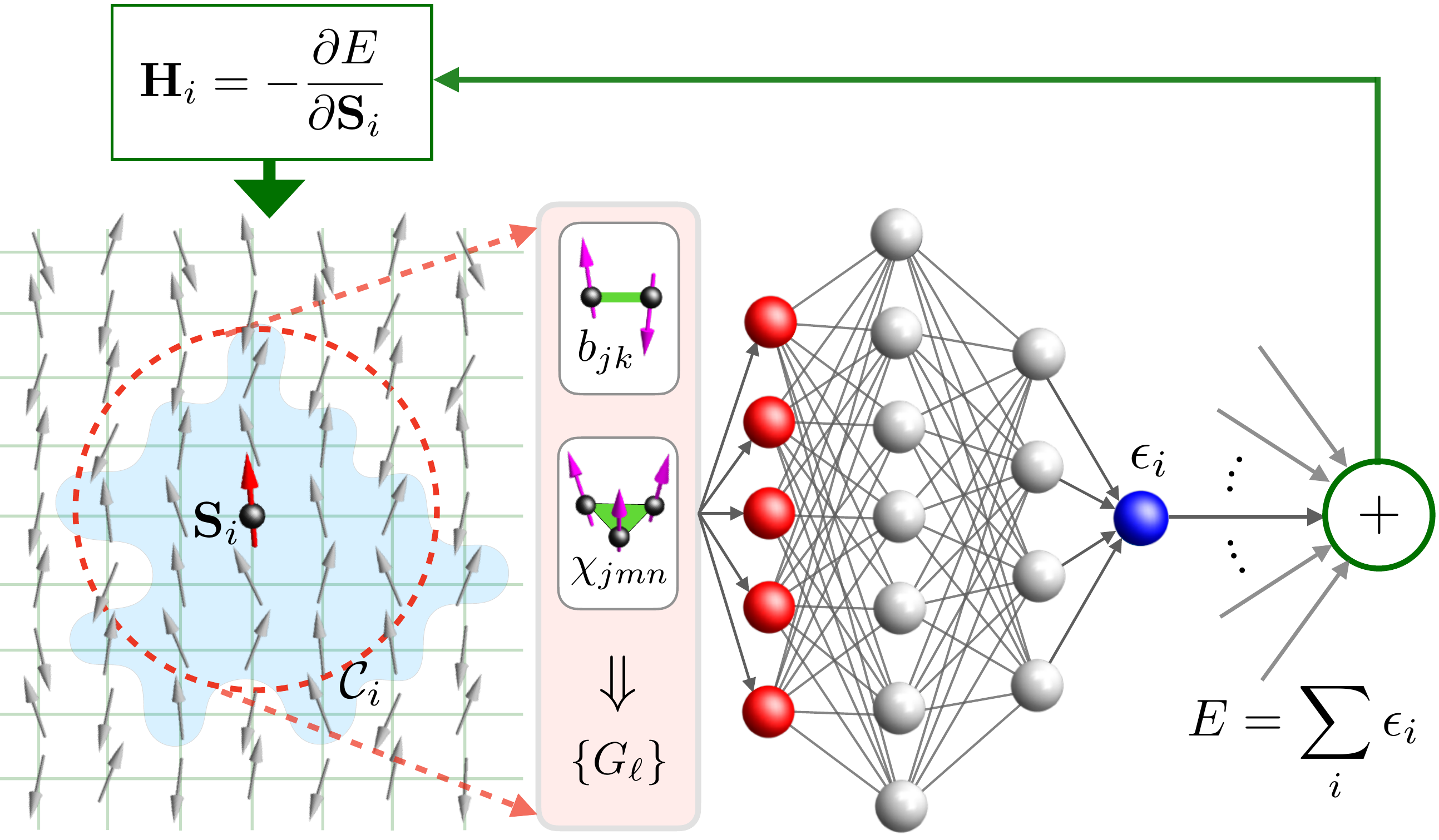}
\caption{Schematic diagram of neural-network (NN) potential model for LLG dynamics simulation of DE system. A descriptor generates the neighborhood spin configuration $\mathcal{C}_i$ to effective coordinates $\{G_\ell\}$ which are then fed into a NN.  The output of the NN is the local energy~$\epsilon_i = \varepsilon(\mathcal{C}_i)$ associated with site-$i$. Automatic differentiation applied to the total energy obtained from all sites gives the local exchange forces~$\mathbf H_i$.
}
\label{fig:ml-scheme}  \
\end{figure}

The evolution of the DE system in the adiabatic limit, similar to the Born-Oppenheimer approximation in quantum molecular dynamics~\cite{marx09}, is governed by the stochastic Landau-Lifshitz-Gilbert (LLG) equation~\cite{landau35,brown63,garanin97}
\begin{eqnarray}
	\label{eq:LLG}
	\frac{d \mathbf S_i}{dt}\ = \mathbf S_i \times \left( \mathbf H_i + \bm\zeta_i \right) - \alpha \mathbf S_i \times \left(\mathbf S_i \times \mathbf H_i \right),
\end{eqnarray}
where $\bm\zeta_i(t)$ is a Gaussian stochastic field of zero mean, $\alpha$~is a damping coefficient, $\mathbf H_i = -\partial E/\partial \mathbf S_i$ is the local exchange force acting on the $i$-th spin. The effective energy $E$ is given by $E =  {\rm Tr}(\hat{\rho} \hat{\mathcal{H}})$, where $\hat{\rho} = \exp\bigl(-\hat{\mathcal{H}} /k_B T\bigr)$ is the instantaneous density matrix of the electron liquid. Repeated calculation of $\rho$, which is required at every time-step, based on exact diagonalization (ED), can be overwhelmingly time consuming~\cite{note_KPM,furukawa04,alvarez05,weisse06,barros13,wang18}. 

To overcome this computational bottleneck, we develop a neural network (NN) model for the potential energy surface $E\left(\{\mathbf S_i\} \right)$ of spins. We first express the effective energy as a sum of local contributions
\begin{eqnarray}
	E = \sum_i \epsilon_i =\sum_i \varepsilon(\mathcal{C}_i),
\end{eqnarray}
where the energy $\epsilon_i = \varepsilon(\mathcal{C}_i)$ is associated with the $i$-th lattice site and is assumed to depend only on spin configuration $\mathcal{C}_i = \{ \mathbf S_j \, | \, r_{ij} < r_c \}$ in its neighborhood. The partitioning of $E$ into local energies is based on the locality principle~\cite{kohn96,prodan05}, which also underlies the ML interatomic potentials that allow for large-scale molecular dynamics simulations with the accuracy of density function theory~\cite{behler07,bartok10,li15,behler16,smith17,zhang18,deringer19,mueller20} or other many-body techniques~\cite{mcgibbon17,chmiela17,suwa19}.

\begin{figure}
\includegraphics[width=0.99\columnwidth]{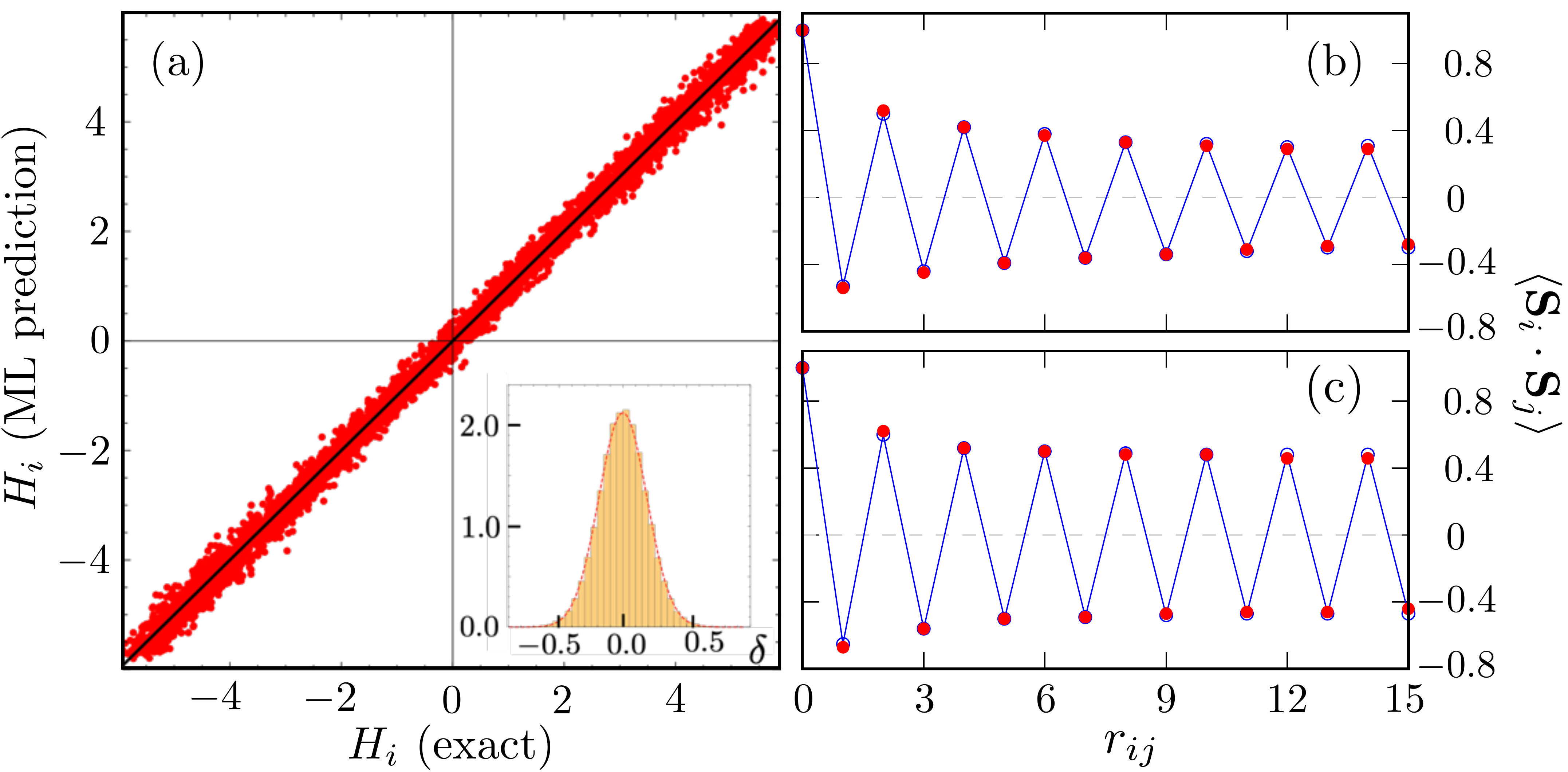}
\caption{(a)  ML predicted exchange forces versus exact solutions from test dataset. The inset shows distribution of the force difference $\delta = H_{\text{ML}}-H_{\text{exact}}$ between ML prediction and ED, which is well approximated by a normal distribution, shown as the red line, with a variance~$\sigma^2 = 0.035$. Right: spin-spin correlation $ \langle \mathbf S_i \cdot \mathbf S_j \rangle$ as a function of $r_{ij} = |\mathbf r_j - \mathbf r_i|$ along the $x$ direction at electron filling fraction (b) $n = 0.485$ and (c) $n = 0.475$.  The red dots are results from LLG simulations with NN models, while the blue dashed lines correspond to ED-LLG simulations at~$T = 0.022$.
\label{fig:force} 
}
\end{figure}

As shown in Fig.~\ref{fig:ml-scheme}, the dependence of energy function $\varepsilon(\mathcal{C}_i)$ on the local spin-environment is encoded in a feed-forward neural network.
To ensure that symmetries of the DE Hamiltonian, which are described by the SO(3) spin-rotation and D$_4$ point groups, are preserved in the energy function, we have developed a descriptor that translates local spin configuration~$\mathcal{C}_i$ into effective coordinates $\{G_\ell\}$ that are invariant under both symmetry operations. First, the SO(3) rotation symmetry can be manifestly maintained by using only bond variables $b_{jk}$ and scalar chirality $\chi_{jmn}$ as building blocks; they are defined as
\begin{eqnarray}
	b_{jk} = \mathbf S_j \cdot \mathbf S_k, \qquad \chi_{jmn} = \mathbf S_j \cdot \mathbf S_m \times \mathbf S_n,
\end{eqnarray}
The collection of these variables around the $i$-th spin $\{b_{jk}, \chi_{jmn} \}$~form the basis of a high-dimensional representation of the D$_4$ group, which is then decomposed into the fundamental irreducible representations (irrep). The basis of each irreps $f^{A_1}_r$, $f^{A_2}_r$, $\cdots$, $\bm f^E_r$, where $r$ enumerates the multiplicity, are proper linear combinations of the bond and scalar chirality variables. Finally, generalized coordinates $\{G_\ell\}$ that are invariant under lattice symmetry operations are obtained from the amplitudes and relative phases of these irrep basis~\cite{ma19,suppl1}. These generalized coordinates are then fed into a NN, which produces the local energy $\epsilon_i$ at its output. Exchange forces $\mathbf H_i$ acting on spins are obtained by applying automatic differentiation to the total energy; see Fig.~\ref{fig:ml-scheme}.

\begin{figure}
\includegraphics[width=0.99\columnwidth]{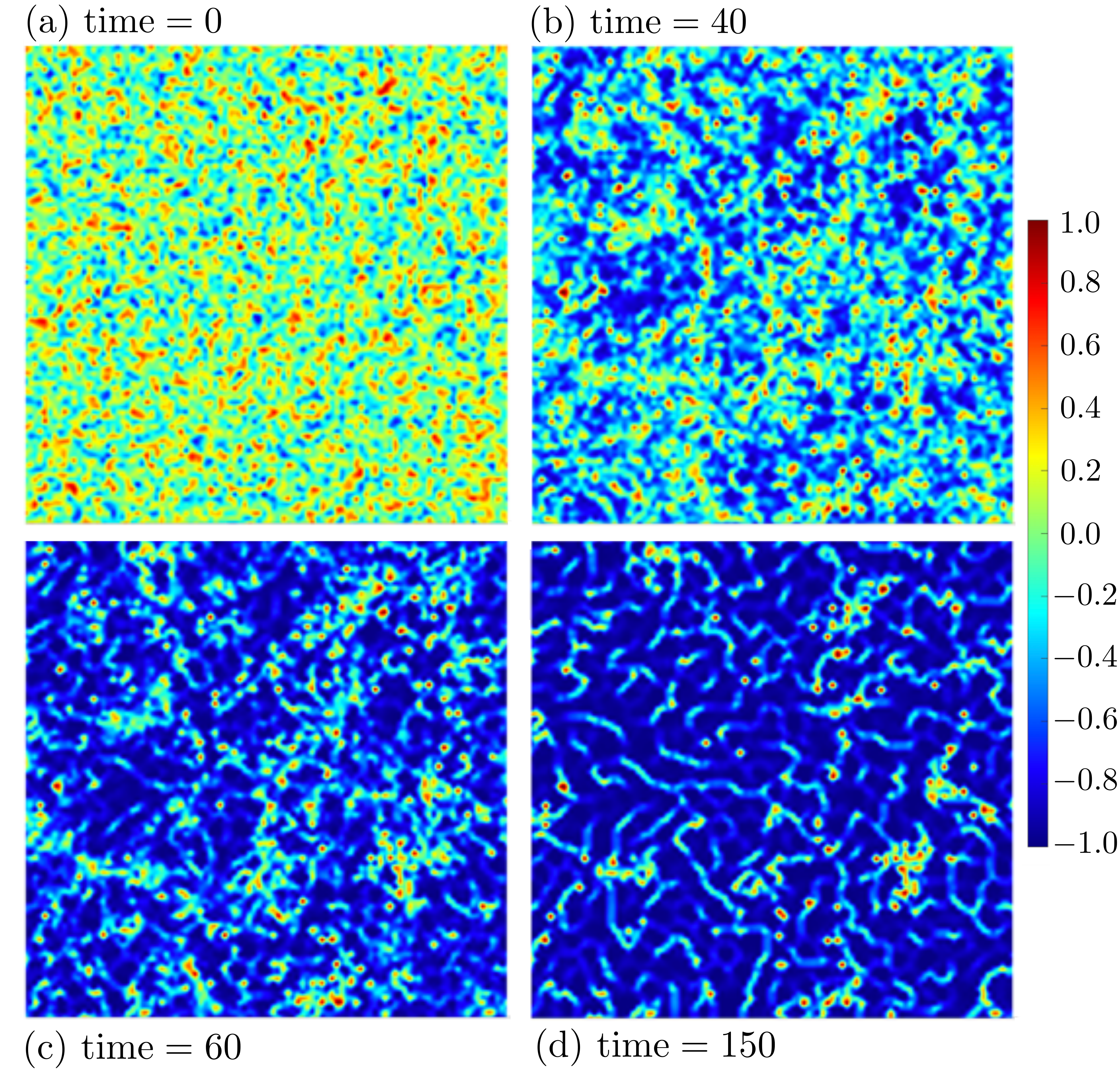}
\caption{Density plots of local bond-variables $b(\mathbf r_i)$ at four different times of the ML-LLG dynamics simulation on a $100 \times 100$ lattice with $1.5\%$ hole doping. The NN model is trained by datasets from exact solutions on a $L=30$ lattice, with parameters $t_{\rm nn} = 1$, $J_H = 7$, and electron filling fraction $n = 0.485$. The simulation time is measured in units of $t_{\rm nn}^{-1}$. 
\label{fig:relaxation} 
}
\end{figure}

A six-layer NN model is constructed and trained using PyTorch~\cite{paszke19}. The training dataset consists of 3500 snapshots of spins and local exchange forces, obtained from exact LLG simulations of a $30\times 30$ lattice~\cite{suppl2}. 
Fig.~\ref{fig:force}(a)~shows components of local exchange forces~$\mathbf H_i$ predicted by our trained NN model versus the exact results on a test dataset consisting of 500 snapshots of spins during the relaxation process. The difference $\delta = H_{{\text {ML}}} - H_{{\text {exact}}}$ between the ML prediction and exact calculation is well described by a Gaussian distribution with a rather small MSE of $\sigma^2 = 0.035$, as shown in the inset. Interestingly, the normal distribution of the deviation $\delta$ implies that the statistical error of the ML model can be interpreted as an effective temperature in the Langevin-type dynamics. This interesting observation has also been verified in our simulations.   As shown in Fig.~\ref{fig:force}(b) and (c), the spin-spin correlations obtained from ML-LLG simulations agree very well with those from ED-based LLG at the temperature $T = 0.022$.

\begin{figure}
\includegraphics[width=0.99\columnwidth]{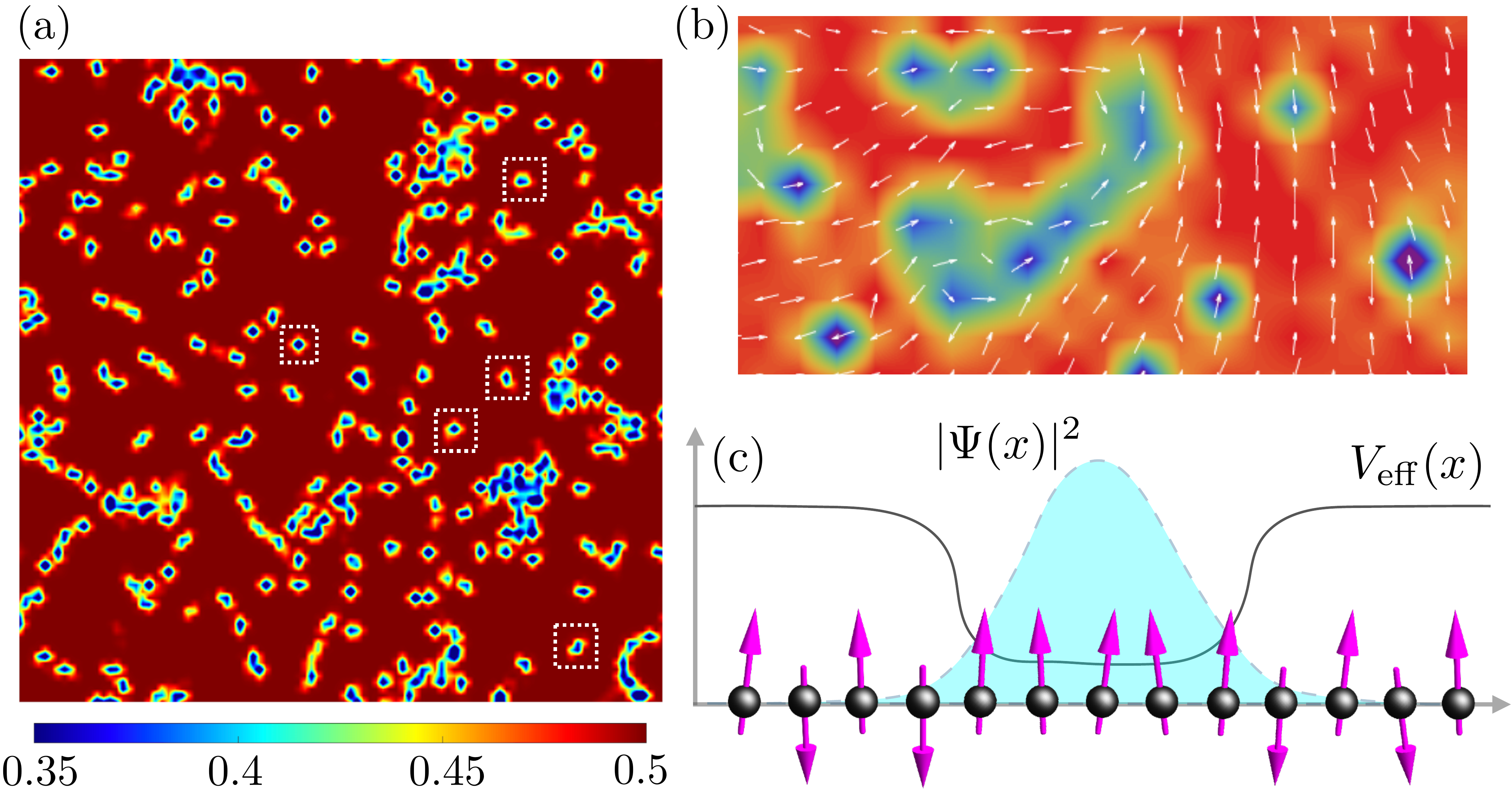}
\caption{(a) Density plot of on-site electron density $n_i$, predicted from NN models for the spin configuration shown in Fig.~\ref{fig:relaxation}(d). Some magnetic polarons are highlighted by dotted squares. (b) A close-up look of the FM clusters and the accompanying spins (projected to a 2D plane). (c) Schematic diagram showing self-confinement of carriers in a FM cluster. 
\label{fig:density} 
}
\end{figure}

Having successfully benchmarked the NN model, we used it to perform large-scale quench simulations in which a $100\times 100$ system, initially in a random configuration, is suddenly quenched to a low-temperature phase. 
Fig.~\ref{fig:relaxation}~shows density plots of local spin-correlation obtained by averaging over 4 nearest-neighbor bonds of a given site, $b_i \equiv (\mathbf S_i \cdot \mathbf S_{i+\mathbf x} + \mathbf S_i \cdot \mathbf S_{i - \mathbf x} + \mathbf S_i \cdot \mathbf S_{i+\mathbf y} + \mathbf S_i \cdot \mathbf S_{i - \mathbf y})/4$, at different times after  quench. Positive~$b_i$ corresponds to regions with predominately ferromagnetic (FM) alignment of spins, while negative $b_i$ indicates antiferromagnetic (AFM) domains. Our ML-LLG simulations clearly showed a relaxation process that leads to an inhomogeneous state with large AFM domains  interspersed with small FM clusters. 
We have trained another NN-model which successfully predicts the on-site electron density $n_i = \frac{1}{2} \sum_{\sigma} \langle \hat{c}^\dagger_{i,\sigma} \hat{c}^{\,}_{i,\sigma} \rangle$ based on the neighborhood spins~$\mathcal{C}_i$. Applying this NN to spin configurations obtained form the ML-LLG quench simulations, we verified that the doped holes are indeed confined to puddles with FM spins, as shown in Fig.~\ref{fig:density}(a) and (b). Interestingly, compared with the electron density, the spins exhibit more complex structures. In particular, in addition to FM clusters, a web of string-like features can be seen in the AFM background of the phase-separated states; see e.g. Fig.~\ref{fig:relaxation}(d).





\begin{figure}
\includegraphics[width=0.98\columnwidth]{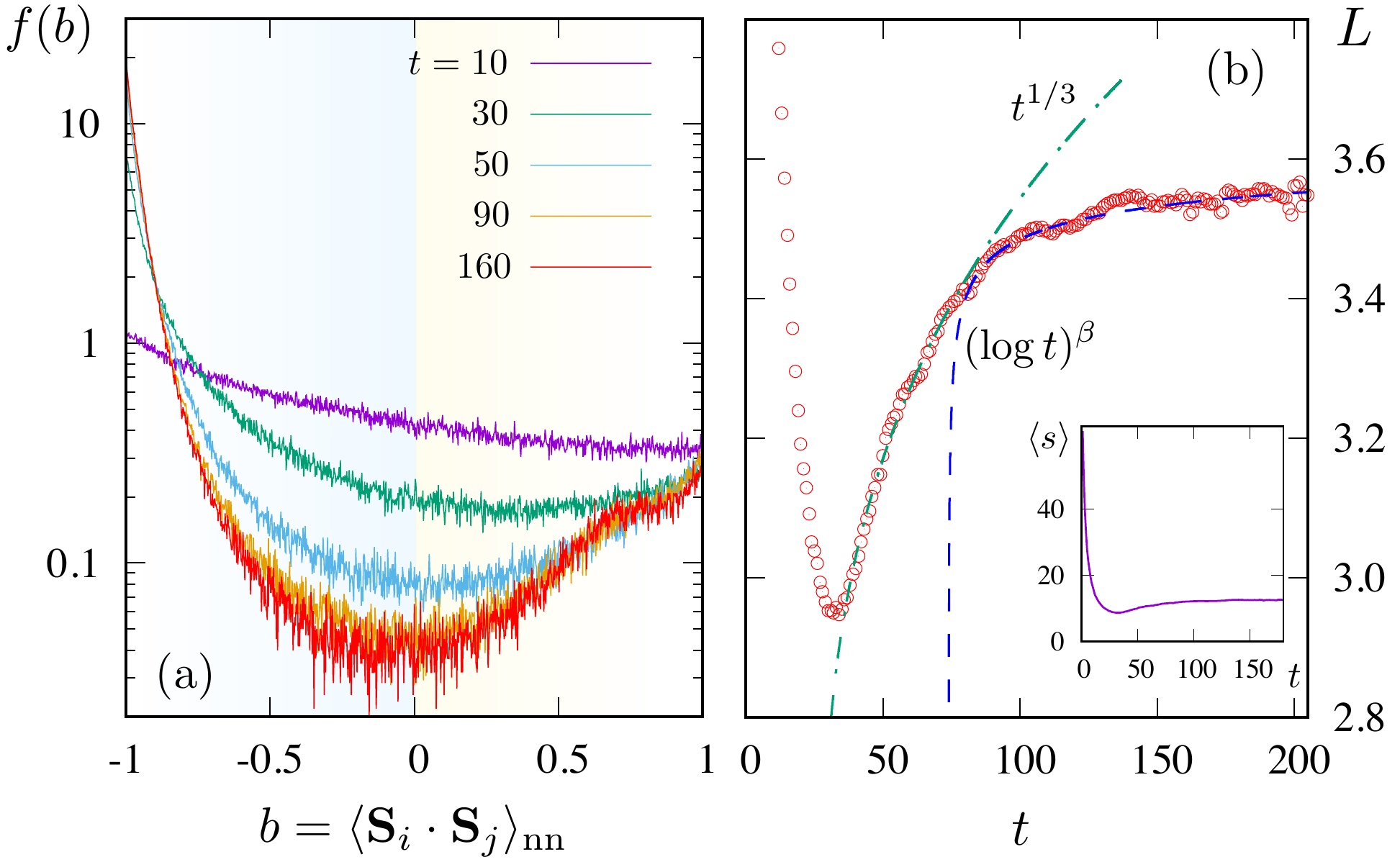}
\caption{(a) Distribution function of the nearest-neighbor bond-variables at different dimes during phase separation. (b)~Average linear size $L = \langle s \rangle^{1/2}$ of FM clusters versus time after a thermal quench. The dash-dot line shows the $t^{1/3}$ power-law growth, while the  dashed line indicates sublogarithmic dependence $L(t) \sim (\log t)^{\beta}$ with $\beta = 0.11$. The inset shows the time dependence of average size $\langle s \rangle$ of FM clusters. 
\label{fig:dynamics} 
}
\end{figure}

Next we turn to the kinetics of FM-domain growth. Fig.~\ref{fig:dynamics}(a) shows the distribution of nearest-neighbor bond-variable $b = \langle \mathbf S_i \cdot \mathbf S_j \rangle_{\rm nn}$ at different times after quench. The initially flat distribution function  $f(b) = 1/(b_{\rm max} - b_{\rm min} ) = \frac{1}{2}$, corresponding to random spins, starts to develop a peak at $b_{\rm min} = -1$ representing AFM spin correlation at earlier times (e.g. at $t = 10$).  This then turns into a bimodal distribution at late times, clearly indicating the evolution of the system toward phase separation, although the peak at the FM-bond $b_{\rm max} = +1$ is rather weaker. Since doped holes in the phase-separated states are mostly confined in FM-clusters, the smaller value of $f(b_{\rm max})$  is consistent with the small doping of our system. 

We define a FM cluster in such a way that all nearest-neighbor bonds within it are greater than a threshold $b_{\rm th} = 0.5$. The time dependence of the characteristic length $L$ of such FM clusters is shown in Fig.~\ref{fig:dynamics}(b).  Qualitatively similar behaviors were obtained using larger threshold~$b_{\rm th}$. In the initial state with random spins, a fraction $\frac{1}{2}(1-b_{\rm th}) = \frac{1}{4}$ of bonds are above the threshold. Although this fraction is still below the bond-percolation threshold $p_{\rm th} = 1/2$ on square lattice, relatively large FM clusters can still be found in the random spins, which explains the initially large average size $\langle s \rangle$ of such FM clusters.  As the system relaxes toward equilibrium, the average cluster size quickly decreases as shown in the inset of Fig.~\ref{fig:dynamics}(b). After reaching a minimum, the hole-rich FM-clusters start to grow again.

Since the number of doped holes is conserved, phenomenologically the growth of such conserved fields is described by the Cahn-Hilliard equation~\cite{cahn58}, also called the model-B dynamics~\cite{hohenberg77}, and a power-law growth $L(t) \sim t^{1/3}$ is expected. Moreover, since the density of doped holes is very small in the mixed-phase states of DE model, the phase-separation dynamics in such off-critical or asymmetric quenches is supposed to be well described by the original LSW theory~\cite{lifshitz61,wagner61}, which convincingly predicts the same $t^{1/3}$ growth. However, as shown in Fig.~\ref{fig:dynamics}(b), only a short initial period of the domain growth can be described by the $\alpha=1/3$ power law. At late times, the length scale $L$ increases with a significantly slower rate than that predicted by the LSW theory, and is better described by a sublogarithmic growth law.

The LSW theory describes the diffusive interactions between domains of  conserved minority phase. Clusters of the minority phase compete for growth through an evaporation-condensation mechanism, whereby larger clusters grow at the expense of smaller ones.  
According to this scenario, growth of the hole-rich FM clusters requires the diffusion of doped holes migrating from smaller cluster to larger ones in the mixed-phase state. The initial aggregation of the charge carriers into proto-FM domains can probably be described by the LSW mechanism, as evidenced by the early-stage power-law growth in our simulations. However, the phase-separation process is dramatically slowed down when the AFM correlation is established in the background of the half-filled majority phase. At this point, the doped holes induce a cloud of surrounding parallel spins through the double-exchange mechanism, which in turn provide a confining potential. Fig.~\ref{fig:density}(b) shows examples of hole-rich FM clusters embedded in a AFM background, and the self-confinement of holes in such domains is schematically shown in Fig.~\ref{fig:density}(c). Importantly, as a result of such self-trapping, evaporation of doped holes from FM clusters is strongly suppressed, leading to the breakdown of the LSW mechanism.

Moreover, even if some charge carriers manage to escape confinement of the emerging FM-clusters described above, they are soon transformed to relatively immobile quasi-particles called magnetic polarons~\cite{degennes60,kasuya68,nagaev74,mauger83,daghofer04,pereira08,jing21}, some of which are highlighted in Fig.~\ref{fig:density}(b). Magnetic polaron of the single-band DE model consists of as few as five parallel spins that trap exactly one fundamental electric charge~\cite{jing21}. In fact, magnetic polarons to some extent can be viewed simply as the smallest FM cluster~\cite{dagotto_book}.

Another mechanism for $L\sim t^{1/3}$ domain-growth, proposed by Binder and Stauffer (BS)~\cite{binder74,binder77}, is based on the Brownian motion and collision of the minority-phase droplets. For the single-band DE Hamiltonian studied in this work, the FM clusters, including magnetic polarons, are pretty much immobile up to temperatures close to the magnetic phase transition~\cite{daghofer04,pereira08,jing21}. This indicates the BS scenario cannot produce a sustained domain growth in our case. On the other hand, taking into account the quantum nature of localized spins, it has been argued that diffusive motion of small magnetic polarons can be achieved through quantum-tunneling~\cite{kemeny75,liu79,ioselevich93} or large paramagnetic fluctuations~\cite{calderon00,wegener02}. For most CMR materials, however, such quantum tunneling is suppressed due to the large magnitude of local spins. At any rate, even with diffusive magnetic-polarons, a consistent treatment of tunneling-induced evaporation of holes is required in order to see whether the LSW scaling might be restored. 

In CMR materials and magnetic semiconductors, the formation of FM clusters and magnetic polarons are accompanied by local lattice distortion and orbital ordering~\cite{dagotto_book}, both of which are expected to further stabilize the composite structure, thus reducing the mobility of charge-carriers. Consequently, similar freezing effect is likely to take place in the phase separation process of real materials. The presence of quenched disorder most likely enhances the glassy behaviors discussed above. For example, it has been shown that charge carriers can be trapped by impurities, forming bound magnetic polarons~\cite{dietl83,mohan88,kaminski03}. Other factors that affect the carrier mobility include hole concentration, external electric and magnetic fields. In particular, higher doping percentage could increase the overlap of the carrier wave function, thus enhancing the tunneling mobility.  

The functionality of correlated electron materials, such as CMR manganites, depends intimately on the structure of the mixed-phase states, which in turn are determined by the phase-separation process. Some of reported anomalous or glassy dynamics in CMR manganites~\cite{yan11,ward11,kundhikanjana15} might be related to the freezing behavior studied in this work.  With powerful ML methods, generalizations to more realistic models which include, e.g. multi-orbital or Jahn-Teller effect, are straightforward. ML-enabled large-scale simulations offer the capability to systematically investigate and characterize phase-separation dynamics, paving the way toward engineering electronic mixed-phase states  with desired properties.

\begin{acknowledgments}
The authors thank Sheng Zhang for useful discussions. The work was supported by the US Department of Energy Basic Energy Sciences under Contract No. DE-SC0020330. The authors also acknowledge the support of Advanced Research Computing Services at the University of Virginia.
\end{acknowledgments}


\begin{thebibliography}{99}
\bibitem{note_KPM} Linear-scaling approaches, such as the kernel polynomial method, have also been developed for the simulating DE systems; see Refs.~\cite{furukawa04,alvarez05,barros13,weisse06,wang18} for examples. The ML-approaches proposed in this work offers more flexibility and can be generalized to include effects such as multi-orbitals, Jahn-Teller coupling, without much overhead.
\end{thebibliography}
\end{document}